\documentstyle[seceq,epsf,wrapft,rotate]{ptptex}
\title{%        
Overview of Neutrino-Nucleus Interactions
}
\author{%       %Use \sc for the family name
A.B. {\sc Balantekin}\footnote{E-Mail: {\tt
    baha@nucth.physics.wisc.edu}} 
}

\inst{%         
Physics Department, University of Wisconsin, Madison, WI  53706  USA}

\recdate{%      %Editorial Office will fill in this.
}

\abst{%       
An overview of neutrino-nucleus
interactions is presented. After a brief discussion of theory and
applications of detector response three examples are discussed: i)
Recent work on neutrino-deuteron reactions, ii) Probing strangeness
content of the nucleon with neutrinos, and iii) The role of neutrino
reactions in supernova dynamics and r-process nucleosynthesis. 
}

\begin{document}

\maketitle

\section{Introduction}

Recent observations of solar
\cite{Fukuda:2001nk,Fukuda:2001nj,Ahmad:2001an} and atmospheric
neutrinos \cite{Fukuda:2000np} at the Super-Kamiokande and Sudbury
Neutrino  Observatories are among the most exciting experimental
results obtained in the last decades. These and other neutrino
observations often utilize neutrino interactions with various
nuclei. Neutrino-nucleus scattering is not only a useful tool for
neutrino detection, but also plays an important role in understanding
the dynamics of a core-collapse supernova and the subsequent
nucleosynthesis. In addition, neutrinos can be used as a probe of
fundamental physics at low energies. The purpose of this talk is to
present a brief overview of neutrino-nucleus interactions to highlight
the rich nuclear physics program that can be carried out at facilities
that can produce low-energy neutrinos. 

One can divide investigation of neutrino-nucleus interactions into 
roughly four groups: 
\begin{itemize}
\item {\bf Theory and Applications of Detector Response:} Solar,
  atmospheric, accelerator, and reactor neutrinos have different
  energies necessitating the calculation of detector response to
  neutrinos at different energy ranges. 

\item {\bf Input into Astrophysics:} This includes a study of  
  neutrino reactions in core-collapse supernovae and also 
  input into the r-process nucleosynthesis calculations. 

\item {\bf Tests of Nuclear Structure Calculations and Approximation
    Methods:} Examples in this category are i) Comparison of indirect
    methods to calculate Gamow-Teller matrix elements (such as (p,n)
    reactions) to possible direct systematic studies and ii) Using
    effective field theories both to directly calculate
    neutrino-deuteron interactions and check for inconsistencies in
    the potential model calculations.

\item {\bf Probing Fundamental Physics at Low Energies:} This can lead
  to a rich physics program. One example is determination
  of the proton strange form factors to better understand the
  strangeness contribution to the proton spin. 
\end{itemize}
In the following sections, after a brief discussion of theory and
applications of detector response, three examples are discussed  to
illustrate the breadth of the phenomena that relate to 
neutrino-nucleus scattering: i) Recent work on neutrino-deuteron
reactions and related tests of effective field theories, ii) Probing
strangeness content of the nucleon with neutrinos, and iii) The role
neutrino interactions in supernova dynamics and r-process
nucleosynthesis.

\section{Detector response}

An elementary introduction to the calculation of neutrino cross
sections is given in Ref. \citen{Balantekin:1999re}, and recent
results are summarized in Refs. \citen{Vogel:1998yz} and
\citen{Haxton:1999nk}.  One can divide studies of detector response
roughly into four categories from lowest to highest neutrino
energies. For solar neutrino energies where one has mostly bound
discrete final states, Shell Model and random phase approximation can
be used. Here the appropriate targets include chlorine
[$^{37}$Cl] \cite{chlorine}\tocite{Adelberger:yd}, gallium [$^{71}$Ga]
\cite{Hampel:1998xg}\tocite{Mathews:tc}, iodine [$^{127}$I]
\cite{Haxton:jh,Engel:sq}, molybdenum [$^{100}$Mo]
\cite{Ejiri:1999rk}, and ytterbium [$^{176}$Yb]
\cite{Raghavan:ad,Fujiwara:2000pr,Bhattacharya:2000ga}. One should
point out that even when solar neutrinos are detected via their
interactions with electrons in a water detector such as
Super-Kamiokande, neutrino interactions with $^{18}$O, present in
water in trace amounts, may be significant \cite{Haxton:1998vj}. 

For low-energy laboratory neutrinos such as those used by the LSND
\cite{Auerbach:2001hz},  KARMEN \cite{Armbruster:gk,Kretschmer:zy},
KamLAND \cite{Piepke:tg}, and Borexino \cite{Alimonti:2000xc}
experiments, one has both bound discrete and continuum final states
and the appropriate targets are carbon [$^{12}$C] 
\cite{Kolbe:xb}\tocite{Hayes:1999ew} and oxygen [$^{16}$O] 
\cite{Haxton:1990ks}\tocite{Haxton:kc}. 

\begin{wrapfigure}{r}{6.6cm}   % r: RIGHT, 6.6cm: WIDTH
          \epsfxsize=6.6cm \centerline{\epsffile{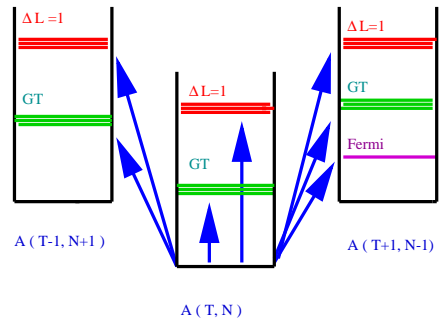}}
          \caption{Interactions of supernova neutrinos.}
          \label{fig:2}
        \end{wrapfigure}               
Supernova neutrino energies are similar to those of low-energy
laboratory neutrinos.  One needs to calculate either capture of
supernova neutrinos on terrestrial 
detectors \cite{McLaughlin:ig}\tocite{Kolbe:2000np} or neutrino
reactions in a supernova environment
\cite{Bruenn:1990qc}\tocite{Toivanen:re} (For a discussion of the role
on neutrinos in the dynamics of core-collapse supernovae and the
r-process nucleosynthesis that can take place in such supernovae
cf. Section 
5). Neutrino interactions that need to be calculated to study a
core-collapse  supernova are depicted in Figure 1. One observes that
not only Fermi and Gamow-Teller transitions, but in a number of cases
also the first-forbidden transitions, need to be calculated. 

Atmospheric neutrinos typically have the highest energies among the
neutrino sources listed here. Neutrino capture in this case is mostly
to the final states in the continuum. 

Collective phenomena are very important in low-energy structure of
nuclei. Since such effects are amenable to algebraic descriptions (see
e.g. Ref. ~\citen{Armia:1976ky}) symmetries significantly simplify
calculations of some nuclear reactions \cite{Balantekin:1997yh}.  One
should point out that collective effects seem not to play an important
role in neutrino capture on nuclei; hence dynamical symmetries of
nuclei do not simplify neutrino capture calculations as they do
calculations of sub-barrier fusion
\cite{Balantekin:1991em,Hagino:1997ze}. It is however possible for
nuclear collectivity to influence other weak interaction processes in
nuclei.  (A discussion of nuclear fusion cross sections that are
important for solar energy generation is beyond the scope of this
talk. For a review see Ref. ~\citen{Adelberger:1998qm}). 

\section{Neutrino-deuteron scattering}

Sudbury Neutrino Observatory is a real-time heavy-water Cerenkov
detector: neutrinos are observed by measuring inelastic
neutrino-deuteron scattering. The only measurement of the total cross
section for charged-current breakup of deuterons by electron neutrinos 
has only about $35\%$ precision \cite{Willis:pj}. Furthermore since
this measurement utilizes decays of stopped muons at the LAMPF
beam-stop the neutrino spectrum resembles that of a supernova, not
that of the Sun. Although the current theoretical accuracy ($\sim 5
\%$) is sufficient to interpret the Sudbury Neutrino Observatory
results, it is important to know neutrino-deuteron cross-sections
better to fully exploit the current and future experimental
results. The calculation of this cross section requires at least three
pieces of input: i) A one-body piece that needs to be convoluted with
the deuteron and two-proton wavefunctions \cite{bahell} (or
two-neutron wavefunction for the electron antineutrino breakup
\cite{Balantekin:dv,Balantekin:1991ns}); ii) A two-body piece which
includes meson-exchange terms \cite{Bahcall:em,Kubodera:rk}; iii)
Radiative corrections \cite{Towner:bh}. Recent calculations that
incorporate modern nucleon-nucleon potentials, but treat
meson-exchange currents with different methods differ by about $5\%$
\cite{Ying:1991tf}\tocite{Kubodera:rk}. 

The calculation of this cross section can be elegantly done using the
tools of effective field theories. Using the power-counting scheme
of Refs. \citen{Kaplan:1998tg} and \citen{Kaplan:1998sz} it was found
that in 
next-to-leading order theoretical uncertainties were dominated by an
unknown axial two-body counter-term \cite{Butler:1999sv} which
eventually needs to be experimentally fixed. Next-to-next-to-leading
order calculations \cite{Butler:2000zp} indicate that such an
expansion indeed converges at neutrino energies of interest. There
is general agreement about the validity and consistency of these
results and the related results for proton-proton fusion
\cite{Park:1998cu}\tocite{Kong:2000px}. Recoil corrections can also be
important as they may effect angular distributions
\cite{Vogel:1999zy}. Finally, a careful analysis of the radiative
corrections indicate that they are of the order of a few percent
\cite{Beacom:2001hr,Kurylov:2001av}.

\section{Strangeness in nuclei}

Neutrino elastic scattering can be used to probe strange quark content
of the nucleon and its contribution to the proton spin. The nucleon 
axial current
\begin{eqnarray}
&{\rm isovector}& \> \> \> {\rm isoscalar}\nonumber\\
\langle N | A^Z_{\mu} | N \rangle  \sim \frac{1}{2} \langle N |
 &\overbrace{ {\overline u} 
 \gamma_{\mu} \gamma_5 u - {\overline d} \gamma_{\mu}
 \gamma_5 d}& - \overbrace{{\overline s} \gamma_{\mu}
 \gamma_5 s} | N \rangle  \nonumber
\end{eqnarray}
has isovector (up and down quarks) and isoscalar (strange
quark) contributions as indicated. The ratio of the
proton-to-neutron knockout reactions, which is dependent on the
isoscalar form factor at $Q^2=0$ (i.e. $\Delta s$), can be used to
study the strangeness contribution \cite{Garvey:qp}:
\[
\frac{ \sigma ( \nu + A \rightarrow p + X)}{\sigma ( \nu + A
  \rightarrow n + X')} \sim  1 +\frac{16}{5} \Delta s .
\]
Extracting $\Delta s$ from such ratios significantly reduces
nuclear-model dependences \cite{Barbaro:1996vd}. 

A second possibility is to look directly for isoscalar excitations,
e.g. 
\[
\nu + ^{12}C \rightarrow ^{12}C^* (12.7 \>{\rm MeV}, 1^+ T=0) + \nu.
\]
Isospin mixing in nuclei certainly complicates such an analysis, but
it may be possible to separate isoscalar and isovector contributions
in certain nuclei \cite{Fujita:zy}. Another possibility to determine
proton strange form factors is neutrino-proton elastic scattering
\cite{Garvey:1992cg}. A recent summary of the studies of strange
quark contributions to nucleon structure is given in
Ref. \citen{Beck:2001dz} and the role of neutrinos is discussed in
Ref. \citen{Holstein:2000pb}. 

\section{Neutrino Interactions in Supernovae}

Understanding neutrino transport in a supernova is an essential part
of understanding supernova dynamics. In a core-collapse  driven
supernova, the inner core  collapses subsonically, but the outer part
of  the core supersonically. At some point during the collapse, when
the nuclear equation of state stiffens, the inner part of the core
bounces, but the outer  core continues falling in. The proto-neutron
star, shrinking under its own  gravity, loses energy by emitting
neutrinos, which only interact weakly and  can leak out on a
relatively long diffusion time scale. Most neutrinos emitted from the
core are produced by a neutral current process, and so the
luminosities are approximately the same for all flavors.  The energy
spectra are approximately Fermi-Dirac with a zero chemical potential
characterized by a neutrinosphere temperature. The $\nu_{\tau},
{\overline \nu}_{\tau}, \nu_{\mu}, {\overline \nu}_{\mu}$ interact
with matter only via neutral current interactions. These decouple at
relatively small radius and end up with somewhat high temperatures,
about 8 MeV. The ${\overline \nu}_e$'s decouple at a larger radius
because of the additional charged current interactions with the
protons, and consequently have a somewhat lower temperature, about 5
MeV. Finally, since they undergo charged current interactions with
more abundant neutrons, $\nu_e$'s decouple at the largest radius and
end up with the lowest temperature, about 3.5 to 4 MeV. Although
numerical values of these temperatures depend on specific models, this
temperature hierarchy is model-independent. 

R-process nucleosynthesis requires a neutron-rich environment, i.e.,
the ratio of electrons to baryons, $Y_e$, should be less than one
half. Time-scale arguments based on meteoritic data suggests that one
possible site for r-process nucleosynthesis is the neutron-rich
material associated with core-collapse supernovae
\cite{Qian:1998,Qian:1998cz}. In one model for neutron-rich material
ejection following the core-collapse, the material is heated with
neutrinos to form a ``neutrino-driven wind''
\cite{Woosley:ux,janka}. In outflow models freeze-out from nuclear
statistical equilibrium leads to the r-process nucleosynthesis. The
outcome of the freeze-out process in turn is determined by the
neutron-to-seed ratio. The neutron to seed
ratio is controlled by three quantities: i) The expansion rate; ii)
The neutron-to-proton ratio (or equivalently the electron fraction,
$Y_e$); iii) The entropy per baryon. Of these three the 
neutron-to-proton ratio is completely determined by the
neutrino-nucleon and neutrino-nucleus interactions. 

Electron fraction in the nucleosynthesis region is given approximately
by \cite{Qian:dg}
\[
Y_e \simeq {1 \over 1+ \lambda_{{\overline \nu}_e p} / \lambda_{ \nu_e
n}}  \simeq {1 \over 1 + T_{{\overline \nu}_e} / T_{ \nu_e}},
\]
where $\lambda_{ \nu_e n}$, etc. are the capture rates and various
neutrino temperatures are indicated by $T$. Hence if $T_{{\overline
\nu}_e} > T_{\nu_e}$, then the medium is neutron-rich. As we discussed
above, without matter-enhanced neutrino oscillations, the neutrino
temperatures satisfy the inequality $T_{ \nu_{\tau}} >T_{{\overline
\nu}_e} > T_{ \nu_e}$. But matter effects via the MSW mechanism
\cite{Balantekin:1998yb}, by heating $\nu_e$ and cooling $\nu_{\tau}$,
can reverse the direction of inequality, making the medium proton-rich
instead. Hence the existence of neutrino mass and mixings puts severe
constraints on heavy-element nucleosynthesis in supernova. These
constraints are investigated in Refs. \citen{Qian:dg} and
\citen{Qian:wh}. One should also point  out that in stochastic media
(i.e. media with large density fluctuations) neutrino flavors would
depolarize \cite{Loreti:1994ry,Balantekin:1996pp}. Although recent
solar neutrino experiments rule out such effects for the Sun
\cite{Balantekin:2001dx}, they may be important in supernovae
\cite{Loreti:1995ae}. 

There are two kinds of neutrino reactions that can destroy r-process:
i) neutrino neutral current spallation of alpha particles \cite{meyer};
ii) formation of too many alpha particles, known as the ``alpha
effect'' \cite{Fuller:ih,Meyer:1998sn}. The alpha effect comes at the
epoch of alpha-particle formation: protons produced by $\nu_e$ capture
on neutrons will in turn capture more neutrons to bind into alpha
particles, reducing the number of free neutrons available to the
r-process and pushing $Y_e$ towards $0.5$. Reducing the $\nu_e$ flux 
will resolve this problem, but we can only do so at a relatively large
radius so that effective neutrino heating already can have
occurred. One way to achieve this is transforming active electron
neutrinos into sterile neutrinos
\cite{McLaughlin:1999pd}\tocite{fetter}. 

For the case of active-sterile mixing with fixed values of neutrino
parameters and matter density, for $Y_e > 1/3$ only electron
neutrinos, and for $Y_e < 1/3$ only electron antineutrinos can undergo
an MSW resonance \cite{McLaughlin:1999pd}. If both electron neutrino
and antineutrino fluxes go through a region of neutrons and protons in
equilibrium (i.e. the reactions $\nu_e + n \rightarrow p + e^-$ and
$\bar{\nu}_e + p \rightarrow n + e^+$ are in steady state equilibrium
with the $\nu_e$ and $\bar\nu_e$ fluxes), then no matter what the
initial $Y_e$ is one may expect that the system will evolve to
a fixed point with $Y_e = 1/3$ ensuring a neutron-rich medium
\cite{Nunokawa:1997ct}. Realistic calculations of the supernova
wind models \cite{McLaughlin:1999pd} do not bear out this assessment
as shown below; although the electron antineutrinos are
converted into sterile species, they are regenerated before the
electron fraction in the wind freezes out. 

Neutrino driven wind models, where the outflow is homologous
(i.e. fluid velocity is proportional to the distance) are
characterized by two parameters: entropy per baryon and the expansion
timescale, $\tau$ (i.e. $r=r_0 e^{t/\tau}$). In Refs.
\citen{McLaughlin:1999pd} and \citen{fetter} we solved the neutrino
evolution equations in matter. In addition we tracked the
thermodynamic and nuclear statistical equilibrium evolution of a mass
element and updated the numbers of neutrons and protons at each time
step directly from the weak capture rates.  This coupling of the
neutrino evolution and self-consistent determination of the abundances
is essential to accurately determine the number of neutrons available
for the r-process. The results are illustrated in Figures 2 and 3 for
expansion timescales of $\tau=0.3$ and $\tau=0.9$ seconds
respectively. One observes that there is a wide range of neutrino
parameters which neutralizes the alpha effect and increases the
neutron-to-seed ratio to produce favorable conditions for r-process
nucleosynthesis.  

\begin{figure}[t]
\parbox{\halftext}{
\rotate[r]{\epsfxsize=4.3cm \centerline{\epsffile{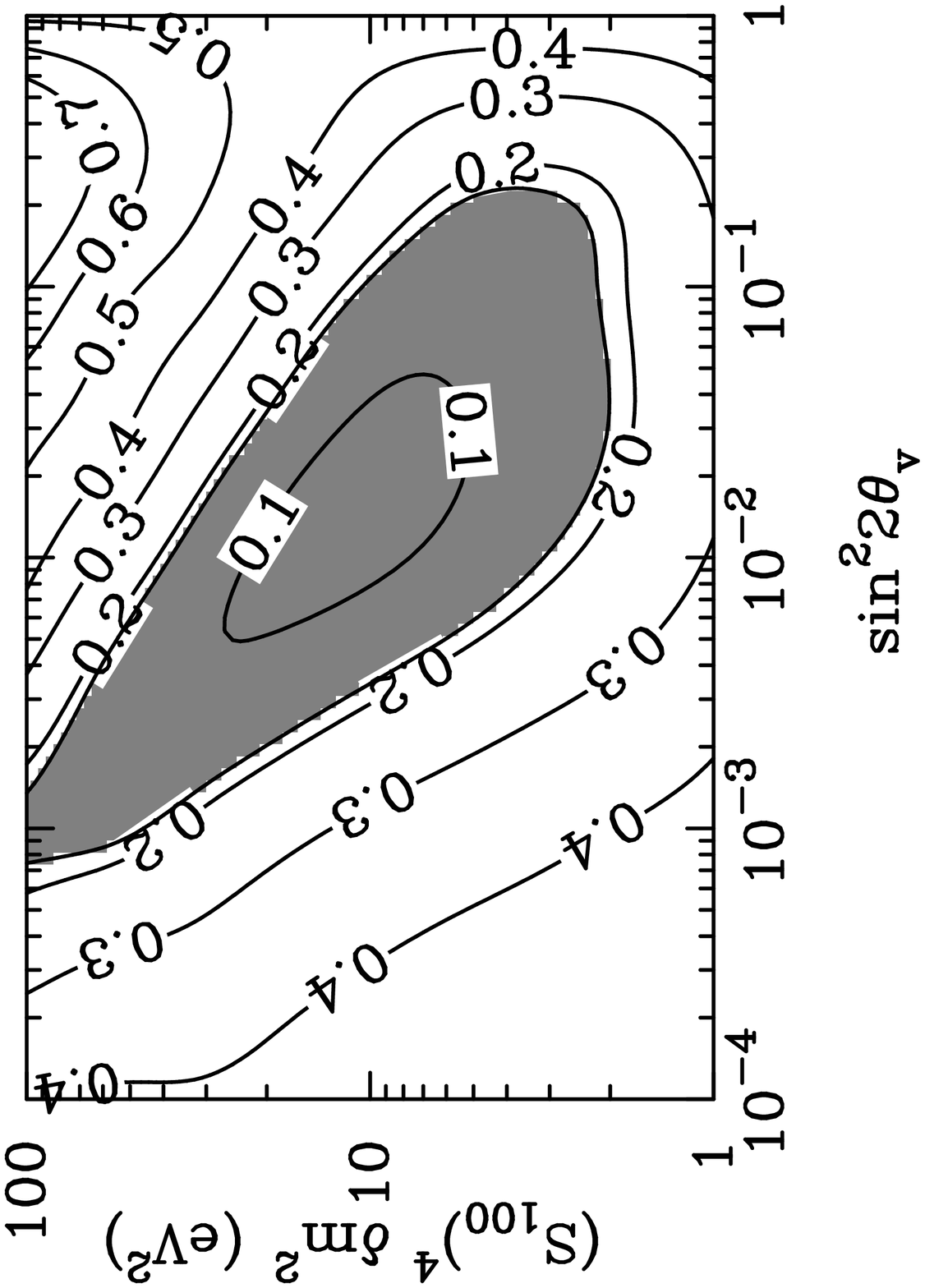}}}
          \caption{Contours of electron fraction for a timescale of
            0.3 s. The shaded area yields a neutron to seed ration of
            at least 100.}}
\hspace{8mm}
\parbox{\halftext}{
\rotate[r]{\epsfxsize=4.3cm \centerline{\epsffile{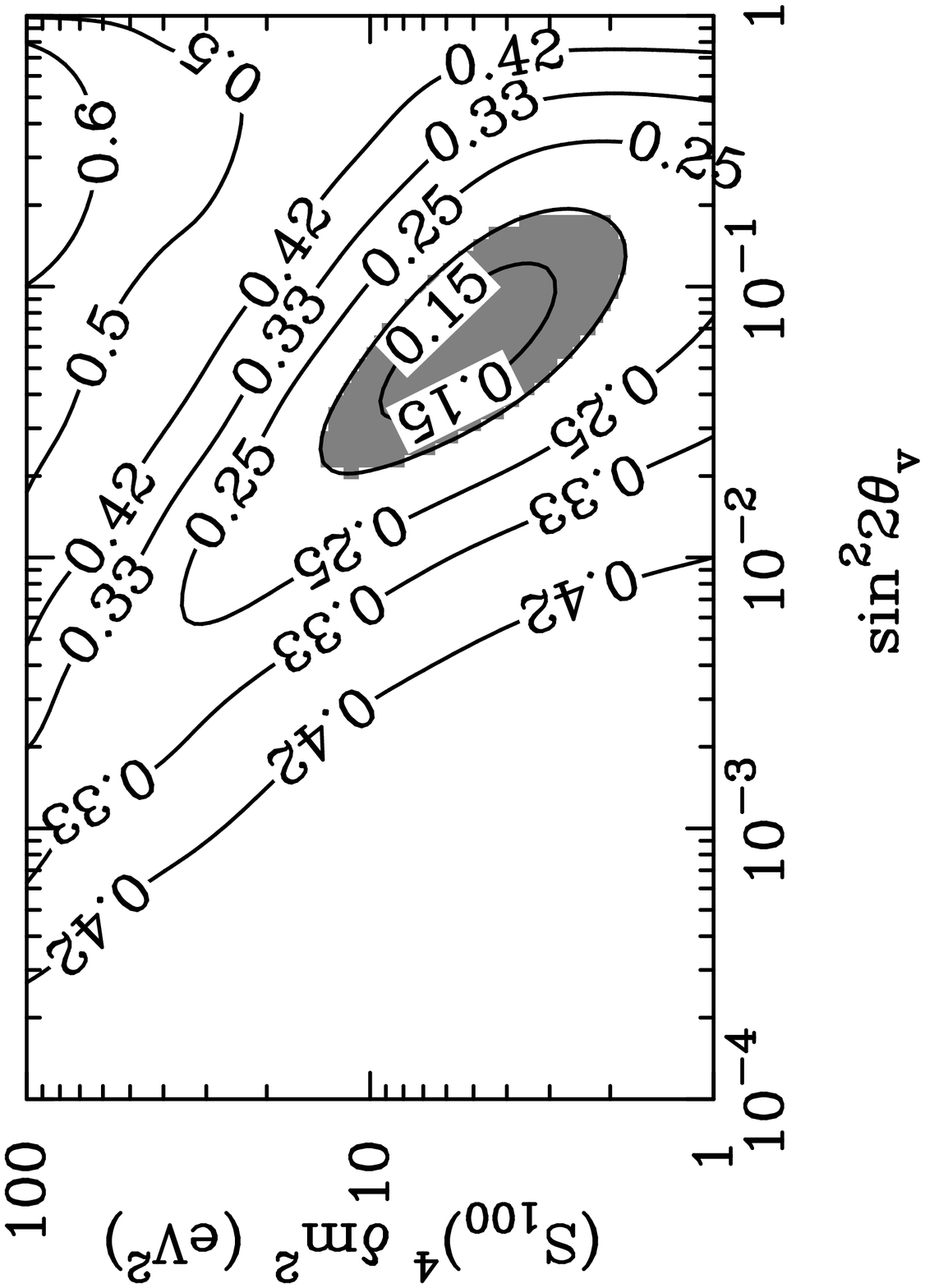}}}
          \caption{Same as Figure 2 for a timescale of 0.9
            seconds. $Y_e \sim 0.5$ in both cases with no flavor
            transformation. }}
\end{figure}

\section{Conclusions}

Various topics covered in this talk were chosen
to illustrate the utility of a comprehensive low-energy nuclear
physics program which could investigate neutrino interactions at
energies appropriate for nuclear and particle astrophysics as well as
nuclear structure. Such a program was originally proposed by the
ORLaND collaboration at Oak Ridge National Laboratory
\cite{Avignone:vh,Avignone:ti}. The Spallation Neutron Source, 
currently under construction at this Laboratory, will produce
neutrinos from the decay of stopped pions created during spallation
and from the decay of the daughter muons. The pulsed nature of such
neutrinos also proves to be useful in eliminating unwanted
backgrounds. The short-term prospects for such a
program seem to be unclear. However, one expects that a
neutrino physics program either at Oak Ridge or at other spallation
neutron facilities being built around the world will eventually be
a reality, since such a program would offer many experimental
advantages at relatively low cost \cite{Bahcall:2000mv}. 

Recent observations of heavy-element abundances in ultra metal-poor
halo stars \cite{metalpoor} provide a deeper insight into the
location of the site for r-process nucleosynthesis. The abundances of
the heavier ($Z>56$) stable neutron-capture elements in such stars
match well the scaled solar system r-process abundances. This concordance 
breaks down for the lighter neutron-capture elements, supporting
previous suggestions \cite{Qian:1998,Qian:1998cz} that different
r-process production sites are responsible for lighter and heavier
neutron-capture elements. Indeed, neutron-star mergers were also
proposed as a site for r-process nucleosynthesis \cite{merger}. 

Figures 2 and 3 indicate that as the expansion timescale gets shorter,
the optimal parameter space moves to larger values of $\Delta
m^2$. However, at very short dynamic expansion timescales the alpha
effect can be very small obviating active-sterile conversion. Indeed
recent calculations indicate that the expansion time scale can be
shorter in some cases \cite{expansion}. A recent comprehensive
overview of element synthesis in stars in given in
Ref. \citen{Thielemann:2001rn}. 

\begin{wrapfigure}{r}{6.6cm}   % r: RIGHT, 6.6cm: WIDTH
          \epsfxsize=6.6cm \centerline{\epsffile{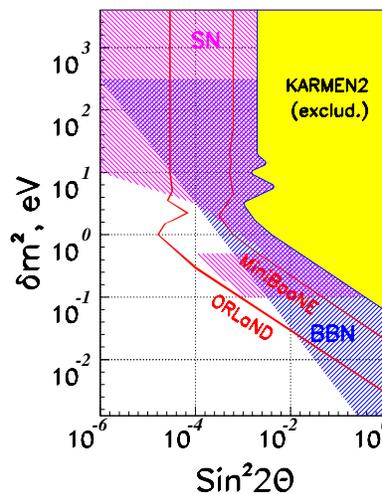}}
          \caption{Regions of neutrino parameter space sensitive to
            supernovae (left-leaning shading) and the early universe
            (right-leaning shading) nucleosynthesis.} 
          \label{fig:4}
        \end{wrapfigure}

In calculating the contours of Figs. 2 and 3 the neutrino mixing
parameter space was freely searched with no reference to the
accelerator neutrino experiments. Nevertheless it is rather
interesting to observe that the optimal parameter space seems to be 
very similar to that which was hinted at by the LSND experiment
\cite{Athanassopoulos:1997pv}. The MiniBooNE experiment currently
under construction at Fermilab \cite{Bazarko:2000id} will either
provide confirming evidence or prove LSND wrong and set more stringent
limits on the neutrino parameters. 

Similarities between the Big Bang during nucleosynthesis and the
neutrino-driven high-entropy ejecta in a core-collapse supernova were
emphasized in the literature \cite{Fuller:jm}. Indeed they are isospin
mirrors of each other; the role played by the protons in the Big Bang
nucleosynthesis is played by the neutrons in the supernova
nucleosynthesis. In both cases the neutron-to-proton ratio is
controlled by the neutrino interactions. The regions of the neutrino
parameter space that affects nucleosynthesis in these environments are
shown in Fig. 4 (Ref. \citen{yuri}). Consequently one can argue
that experiments such as LSND, KARMEN, ORLaND and MiniBooNE that probe
this region of the neutrino parameter space are relevant to
understanding both the Big Bang and supernova nucleosynthesis (see,
e.g. Ref. \citen{Sorel:2001jn}). Future experiments such as 
MiniBooNE or ORLaND (or similar detectors), no matter what their
outcomes are, will probe this remarkable interplay between element
production in the cosmos and the fundamental properties of neutrinos. 

\section*{Acknowledgments}

I thank G. Fuller, B. Holstein, A. Mezzacappa, G. McLaughlin, and
Y.Z. Qian for may useful discussions. 
This work was supported in part by the U.S. National  Science
Foundation Grant No.\ PHY-0070161  at the University of Wisconsin, and
in part by the University  of Wisconsin Research Committee with funds
granted by the Wisconsin Alumni Research Foundation.  I would like to
thank the organizers of the conference for their kind hospitality.

\end{document}